\documentstyle[12pt]{article}
\begin{document}
\title{\vspace{-1cm} \hfill {\small NTUA-47/94}\vspace{-.3cm} \\
\hfill {\small hep-th/9501070 }\vspace{1cm} \\
{\Large {\bf Discontinuities and collision of gravitational waves in string
theory \thanks{%
Partially supported by CEC Science projects No. SC1-CT91-0729 and
SC1-CT92-0792}}}}
\author{{\normalsize Alexandros A. Kehagias} \thanks{%
On leave from Phys. Dept., National Technical Univ. 15773 Zografou Athens,
Greece.} \thanks{%
e-mail: kehagias@sci.kun.nl} \\
{\normalsize Laboratoire de Physique Theorique et Hautes Energies}\\
{\normalsize Universit\'e de Paris-Sud, B\^at. 211, F-91405 Orsay} \\
[.1cm] {\normalsize and} \\
[.1cm] {\normalsize E. Papantonopoulos \thanks{%
e-mail: lpapa@isosun.ariadne-t.gr}} \\
{\normalsize Phys. Dept., National Technical Univ.} \\
{\normalsize 15773 Zografou Athens, Greece}}
\date{}
\maketitle

\begin{abstract}
\begin{sloppypar}
\normalsize
We examine here  discontinuities in the metric,
 the antisymmetric  tensor and the dilaton field
 which  are allowed by conformal invariance.   We find that the surfaces
of discontinuity must necessarily be null surfaces and shock and impulsive
waves are both allowed. We employ our results for
the case of colliding plane gravitational waves and we  discuss
the $SL(2,I\hspace{-.15cm}R)\!\times\!SU(2)/
I\hspace{-.15cm}R\!\times\!I\hspace{-.15cm}R$
 WZW model
 in the present perspective.
In particular, the singularities encountered in this model may be viewed
as the result of the mutual focusing of the colliding waves.
\end{sloppypar}
\end{abstract}

\addtolength{\baselineskip}{.3\baselineskip}

\newpage

An interesting class of solutions in general relativity possessing
remarkable properties consists of plane gravitational waves \cite{1}. These
solutions constitute their own linearized approximation and thus may be
viewed as the exact classical analogs of gravitons. The space-time of such
waves is complete and it does not possess global Cauchy surfaces \cite{1},%
\cite{2}. These solutions can be extended to exact string
solutions \cite{3}.
 Exact plane wave string backgrounds have been obtained by employing WZW
models based on non-semisimple groups as well as various gaugings of the
latters \cite{4}. These models can also been obtained from semisimple WZW
models by taking special singular limits \cite{2},\cite{5}.

On the other hand, waves can, in principle, collide and one may ask if it is
possible a particular string background to be interpreted as the result of
the collision of plane waves, at least in a semiclassical approximation.
Before trying to answer this question, let us recall some known results.

The existence of wave-like solutions in Einstein gravity it has long been
recognized. There exist mainly two types of waves, shock and impulsive ones
\cite{2}. Sock waves have discontinuities in the Riemann tensor, ($C^1$%
--metric), while impulsive ones have a $\delta $-function profile in the
curvature ($C^0$--metric). Shock wave backgrounds have also been discussed
in string theory \cite{6} as well as in field theory because they play an
important role in scattering process in ultra-high energies \cite{7},\cite
{7'}. On the other hand, impulsive waves may be considered as the most
``elementary'' ones and they have been studied in connection with the motion
of massive particles along the horizon of a black-hole background \cite{8},%
\cite{8'}.

However, due to the non-linearity of the field equations, wave solutions
cannot superposed except in the weak field limit. In fact, nowhere the
non-linear character of Einstein gravity shows up more clearly than in the
collision of gravitational waves \cite{9}. Unlike a linear theory, i.e.,
classical electrodynamics, where waves pass straight through each other, in
general relativity waves necessarily tend to focus. For plane waves the
focus usually appears as a singularity in space-time \cite{9}--\cite{11}.
This seems to be a generic feature of the collision provided that the waves
are sufficiently strong to produce serious focusing.

We will discuss below the necessary conditions to have discontinuities in
the target space in string theory in such a way that their presence do not
affect the beta-function equations. We will see that the discontinuities
must be across null hypersurfaces and thus backgrounds resulting from linear
superposition of independent shock or impulsive waves with parallel
propagating wave-fronts are exact as well. We will also discuss the case of
the collision of opposite moving plane waves. For this case, we consider the
$SL(2,I\hspace{-.15cm}R)\!\times \!SU(2)/I\hspace{-.15cm}R\!\times \!I%
\hspace{-.15cm}R$ WZW model \cite{12} and we show that it can be
interpreted, to leading order, as the resulting space-time of the collision
of two such waves.

String propagation in a non-trivial background is described by the 2D $%
\sigma $-model action
\begin{equation}
S=\frac 1{4\pi \alpha ^{\prime }}\int dzd\bar z\left( (G_{\mu \nu
}(X)+B_{\mu \nu }(X))\partial X^\mu \bar \partial X^\nu +\alpha ^{\prime
}R^{(2)}\Phi (X)\right) ,
\end{equation}
where $G_{\mu \nu }(X),\,B_{\mu \nu }(X)$ and $\Phi (X)$ are the target
metric, the antisymmetric tensor and the dilaton field, respectively. At
one-loop level of the coupling constant $\alpha ^{\prime }$, conformal
invariance requires
\begin{eqnarray}
R_{\mu\nu}-\frac{1}{4}H_{\mu\kappa\lambda}{H_{\nu}}^{\kappa\lambda}
+2\nabla_{\mu}\nabla_{\nu}\Phi&=&0,\label{e1}\\
\nabla_\mu(e^{-2\Phi}H^{\mu\nu\rho})&=&0,\label{e2}\\
\frac{2\delta c}{3}-R+\frac{1}{12}H^2-4\nabla^2\Phi+4(\nabla\Phi)^2
&=&0, \label{e3}
\end{eqnarray}
where $H_{\mu \nu \rho }=\nabla _{[\mu }B_{\nu \rho ]}$ is the field
strength of the antisymmetric field and $\delta c$ is the central charge
deficit. One assumes that the vacuum is of the form $M^4\!\times \!K$ where $%
M^4$ is the Minkowski space-time represented by a free theory with $c=4$ and
$K$ is the internal space corresponding to a conformal field theory of
appropriate central charge. By replacing $M^4$ by another four-dimensional
target space N representing again a $c=4$ conformal field theory, one may
obtain other backgrounds for consistent string propagation. Such backgrounds
satisfy Eqs. (\ref{e1}--\ref{e3}) with $\delta c=0$ and may be realized, for
example, as gravitational waves or cosmological solutions. All possible
spaces $N$ considered so far have been assumed to be endowed with metric,
antisymmetric tensor and dilaton fields which are continuous everywhere and
have continuous derivatives as well. Here, we will discuss the possibility
of solving Eqs. (\ref{e1}--\ref{e3}) for four-dimensional space-times $N$ in
which we will allow discontinuities of the metric, antisymmetric tensor and
dilaton field across appropriate hypersurfaces.

To begin with, let us consider the antisymmetric three-form field $%
H_{\mu\nu\rho}$ in $N$ and let us suppose that it has a finite discontinuity
across a hypersurface $\Sigma$ defined by the equation $u(X^\mu)=0$. We may
represent such a field in terms of distributions over a suitable set of test
functions as
\begin{equation}
\label{disH}H_{\mu\nu\rho}=H^{(0)}_{\mu\nu\rho}+h_{\mu\nu\rho}\theta(u),
\end{equation}
where $H^{(0)}_{\mu\nu\rho}$ and $h_{\mu\nu\rho}$ are $C^0$ and piecewise $%
C^1$ and $\theta(u)$ is the Heaviside step function distribution. Thus, we
have
\begin{eqnarray}
H_{\mu\nu\rho}=H^{(0)}_{\mu\nu\rho}&,& u<0,\nonumber \\
H_{\mu\nu\rho}=H^{(0)}_{\mu\nu\rho}+h_{\mu\nu\rho} &,& u>0. \nonumber
\end{eqnarray}
By using the notation
$$
[f]=f^+-f^-,
$$
where $f^+(f^-)$ is the limit of $f$ as one approaches the surface $\Sigma$
from the left (right) $u<0 (u>0)$, we may express Eq. (\ref{disH}) as
$$
[H_{\mu\nu\rho}]=h_{\mu\nu\rho} .%
$$

The closeness of $H_{\mu\nu\rho}$ and Eq. (\ref{e2}) gives
\begin{eqnarray}
\partial_{[\kappa}H^{(0)}_{\mu\nu\rho]}+
\partial_{[\kappa}h_{\mu\nu\rho]}\theta(u)+h_{[\mu\nu\rho}u_{\kappa]}
\delta(u)&=&0 , \nonumber \\
\nabla^{\mu}(e^{-2\Phi}H_{\mu\nu\rho})
+ \nabla^{\mu}(e^{-2\Phi}h_{\mu\nu\rho}) \theta(u)+
e^{-2\Phi}h_{\mu\nu\rho}u^{\mu}
\delta(u)&=&0, \nonumber
\end{eqnarray}
where $u_\mu=\partial u/\partial x^{\mu}$ and $\delta(u)$ is the Dirac $%
\delta$-function. In order these equations to hold, the conditions
\begin{eqnarray}
h_{[\mu\nu\rho}u_{\kappa]}&=&0, \label{a}\\
 h_{\mu\nu\rho}u^{\mu}&=&0. \label{a1}
\end{eqnarray}
have to be satisfied. We may express $h_{\mu\nu\rho}$ as the dual of a
vector $h^{\mu}$
$$
h_{\mu\nu\rho}=\epsilon_{\mu\nu\rho\lambda} h^{\lambda},
$$
($\epsilon_{0123}=+1$) and, consequently, Eqs. (\ref{a},\ref{a1}) are
written as
\begin{eqnarray}
h^{\mu}u_{\mu}&=&0,\nonumber \\
h^{\lambda}\epsilon_{[\lambda\mu\nu\rho} u_{\kappa]}&=&0. \nonumber
\end{eqnarray}
The solution to the above equations is
\begin{eqnarray}
h^{\mu}&=&h u^{\mu}, \nonumber \\
u^{\mu}u_{\mu}&=&0,
\end{eqnarray}
where $h$ is a scalar. As a result, the surfaces of discontinuity must be
null surfaces and the possible discontinuities of $H_{\mu\nu\rho}$ are of
the form
\begin{equation}
\label{H}h_{\mu\nu\rho}=h\epsilon_{\mu\nu\rho\lambda} u^{\lambda}.
\end{equation}

A discontinuity of the type (\ref{disH}) in the $H$-field could only be
emerged from a corresponding one in the antisymmetric tensor of the form
\begin{equation}
B_{\mu\nu}=B^{(0)}_{\mu\nu}+b_{\mu\nu}\theta(u).
\end{equation}
The field strength is then given by
\begin{equation}
\label{hh}H_{\mu\nu\rho}=H^{(0)}_{\mu\nu\rho}+\partial_{[\mu}b_{\nu\rho]}
\theta(u)+b_{[\mu\nu}u_{\rho]}\delta(u),
\end{equation}
and by comparing Eqs. (\ref{disH},\ref{hh}), we find that
\begin{eqnarray}
h_{\mu\nu\rho}&=&\partial_{[\mu}b_{\nu\rho]}
, \\
b_{[\mu\nu}u_{\rho]}&=&0.
\end{eqnarray}
Consequently, the discontinuities in the antisymmetric tensor is of the form
\begin{equation}
b_{\mu\nu}=b_{[\mu}u_{\nu]} .
\end{equation}

Let us now examine the type of discontinuities of the dilaton field $\Phi
(X) $. We will assume that $\Phi (X)$ is continuous while its derivatives
may have finite jumps across the hypersurface $\Sigma $ of the form
\begin{equation}
\label{f}\partial _\mu \Phi =\partial _\mu \Phi ^{(0)}+\phi _\mu \theta (u).
\end{equation}
To find the explicit form of the discontinuity $\phi _\mu $, we expand the
dilaton $\Phi (x)$ in the neighborhood of $\Sigma $ as
\begin{eqnarray}
\Phi(x^+)&=&\Phi_0+\Phi_+^{\prime}u+\frac{1}{2}\Phi_+^{\prime\prime}u^2
+\cdots ,
\nonumber \\
\Phi(x^-)&=&\Phi_0+\Phi_-^{\prime}u+\frac{1}{2}\Phi_-^{\prime\prime}u^2
+\cdots , \label{expan}
\end{eqnarray}
where primes $(^{\prime })$ denote derivatives with respect to $u$ on $%
\Sigma $. By differentiating the above expressions, we find that $\phi _\mu $
is proportional to $u_\mu $, i.e.,
$$
\phi _\mu =\phi u_\mu .
$$
It is easy then to check that this type of discontinuity is compatible with
Eq. (\ref{e3}) since $\delta $-terms coming from differentiation of (\ref{f}%
) disappear because of the nullity of $u_\mu $.

Let us now turn to Eq. (\ref{e1}). One may expect that the appropriate
conditions across surfaces are the Lichnerowicz ones. The latter postulate
continuation of the metric and its first derivatives, i.e., the metric is
considered to be $C^1$ and piecewise $C^2$. Consequently, the Riemann tensor
is piecewise $C^0$ and it allows ``shock'' discontinuities. We may, however,
relax these conditions by considering piecewise $C^1$ metric. In this case
we may write
\begin{equation}
\label{dism}\partial _\rho G_{\mu \nu }=\partial _\rho G_{\mu \nu
}^{(0)}+\gamma _{\mu \nu }u_\rho \theta (u).
\end{equation}
This expression may be obtained by expanding the metric $G_{\mu \nu }$ in
the neighborhood of $\Sigma $ (as in Eq. (\ref{expan})) as
\begin{eqnarray}
G_{\mu\nu}(x^+)&=&G_{\mu\nu0}+G_{\mu\nu+}^{\prime}u+
\frac{1}{2}G_{\mu\nu+}^{\prime\prime}u^2
+\cdots , \nonumber \\
G_{\mu\nu}(x^-)&=&G_{\mu\nu0}+G_{\mu\nu-}^{\prime}u+
\frac{1}{2}G_{\mu\nu-}^{\prime\prime}u^2
+\cdots , \label{expan1}
\end{eqnarray}
The Ricci tensor and the curvature scalar are easily found to be
\begin{eqnarray}
R_{\mu\nu}&=&R^{(0)}_{\mu\nu}+
\frac{1}{2}
(u^{\kappa}u_{\nu}\gamma_{\mu\kappa}
-u_{\mu}u_{\nu}{\gamma_{\kappa}}^{\kappa}-u^{\kappa}u_{\kappa}\gamma_{\mu\nu}
+u_{\mu}u^{\kappa}\gamma_{\nu\kappa})\delta(u), \\
R&=&R^{(0)}+
(u^{\mu}u^{\nu}\gamma_{\mu\nu}-u^{\mu}u_{\mu}{\gamma_{\nu}}^{\nu})
\delta(u),
\end{eqnarray}
with $R_{\mu \nu }^{(0)}$, $R^{(0)}$ piecewise $C^0$. Thus, we way write Eq.
(\ref{e1}) as
\begin{eqnarray}
R^{(0)}_{\mu\nu}-\frac{1}{4}H_{\mu\nu\rho}H^{\mu\nu\rho}+2\nabla_\mu
\nabla_\nu\Phi^{(0)}+2\nabla_\mu(\phi u_\nu)\theta(u)
+\nonumber \\
\frac{1}{2}
(u^{\kappa}u_{\nu}\gamma_{\mu\kappa}
-u_{\mu}u_{\nu}{\gamma_{\kappa}}^{\kappa}-u^{\kappa}u_{\kappa}\gamma_{\mu\nu}
+u_{\mu}u^{\kappa}\gamma_{\nu\kappa})\delta(u)+2\phi u_\mu u_\nu\delta(u)=0.
\end{eqnarray}
and Eq. (\ref{e1}) has been split into a piecewise $C^0$ part and a singular
part. The only way the above equation to hold is
\begin{eqnarray}
R^{(0)}_{\mu\nu}-\frac{1}{4}H_{\mu\nu\rho}H^{\mu\nu\rho}+2\nabla_\mu
\nabla_\nu\Phi^{(0)}+2\nabla_\mu(\phi u_\nu)\theta(u)
&=0&, \label{c0}        \\
u^{\kappa}u_{\nu}\gamma_{\mu\kappa}
-u_{\mu}u_{\nu}{\gamma_{\kappa}}^{\kappa}-u^{\kappa}u_{\kappa}\gamma_{\mu\nu}
+u_{\mu}u^{\kappa}\gamma_{\nu\kappa}&=&-4 \phi u_\mu u_\nu.
\label{cond}
\end{eqnarray}
Eq. (\ref{c0}) can be solved separately in the two regions $u>0,\,u<0$,
while Eq. (\ref{cond}) constrains the possible discontinuities of the
metric. In fact, the solution to the latter equation is provided by
\begin{equation}
\label{BS}\gamma _{\mu \nu }u^\nu =\frac 12{\gamma _\nu }^\nu u_\mu -2\phi
u_\mu .
\end{equation}
Thus, shock waves ($\gamma _{\mu \nu }=0,\,\phi =0$) as well as impulsive
ones ($\gamma _{\mu \nu }\neq 0,\,\phi \neq 0$) are both allowed.

Let us now apply the previous results in the case of the collision of plane
waves. We will consider only ``head on" collision since by making a Lorentz
transformation one can arrange the waves to propagate in opposite spatial
directions. The metric of a plane wave (in harmonic coordinates) is
$$
ds^2= U(u,x,y)du^2-2dudv+dx^2+dy^2 ,%
$$
where
$$
U(u,x,y)=f(u)(x^2-y^2)+g(u)xy.
$$
By performing a suitable coordinate transformation, we may write the metric
above in the form (Rosen coordinate system)
\begin{equation}
ds^2=2e^{-M(u)}dudv+g_{ij}(u)dx^i dx^j,
\end{equation}
where $i,j=1,2$. If the metric $g_{ij}$ of the surface $u=const.$, $v=const.$
is diagonalizable by a linear in $x^i$ transformation, the plane wave has
constant polarization (i.e., $g(u)=0$) and this is the type of waves we are
dealing with.

Let us now assume that space-time admits a two-parameter abelian group of
space-like isometries. The metric for such a space-time can be written as
\begin{equation}
\label{met}ds^2=2e^{-M}dudv+g_{ij}dx^idx^j,
\end{equation}
where
\begin{eqnarray}
M=M(u,v)&,& g_{ij}=g_{ij}(u,v).\nonumber
\end{eqnarray}
We may divide space-time into four distinct region labeled as $%
I:(u<0,\,v<0),\,II:(u>0,\,v<0),\,III:(u<0,\,v>0),\,IV:(u>0,\,v>0)$.
Moreover, we will assume that two plane waves in regions $II$ and $III$
moving in opposite spatial directions approach each other in the flat
Minkowski region $I$ and, subsequently, they collide in region $IV$. Thus,
the metric in this coordinate system will be of the form (\ref{met}) with
\begin{eqnarray}
I:& M=0, & g_{ij}=\delta_{ij},\nonumber \\
II:& M=M(u), &g_{ij}=g_{ij}(u),\nonumber \\
III: &M=M(v), & g_{ij}=g_{ij}(v),\nonumber \\
IV: & M=M(u,v), &g_{ij}=g_{ij}(u,v).
\end{eqnarray}
The metric in region $IV$ is uniquely determined by a characteristic initial
value problem with data determined on the null hypersurfaces $u=0$ and $v=0.$

We consider the six-dimensional $SL(2,I\hspace{-.15cm}R)_{k^{\prime
}}\!\times \!SU(2)_k$ at levels $(k^{\prime },k)$ WZW model with central
charge
$$
c=\frac{3k^{\prime }}{k^{\prime }+2}+\frac{3k}{k+2}
$$
By choosing $k^{\prime }=-k$, we have $\delta c={\cal O}(1/k^2)$ and the
corresponding $\sigma $-model will satisfy Eqs. (\ref{e1}--\ref{e3}) with $%
\delta c=0$. By gauging an anomaly free two-dimensional abelian subgroup $H$
of $SL(2,I\hspace{-.15cm}R)\!\times \!SU(2)$, one obtains a four-dimensional
gauged WZW model with $c=4+{\cal O}(1/k^2)$ and the target space of the
corresponding $\sigma $-model could replace Minkowski space-time. One may
choose to gauge the $H=I\hspace{-.15cm}R\!\times \!I\hspace{-.15cm}R$
subgroup which transforms the elements $(g_1,g_2)\!\in \!SL(2,I%
\hspace{-.15cm}R)\!\times \!SU(2)$ as
\begin{eqnarray}
g_1&\rightarrow &
\exp(\epsilon\sigma_3)g_1\exp(\bar{\epsilon}\cos\alpha\sigma_3
+\epsilon\sin\alpha \sigma_3),\label{h}\\
g_2&\rightarrow &
\exp(\bar{\epsilon}\sigma_3)g_2\exp(i\epsilon\cos\alpha\sigma_2
-i\bar{\epsilon}\sin\alpha \sigma_2),\label{h1}
\end{eqnarray}
where $(\sigma _i,\,i=1,2,3)$ are the standard Pauli matrices and $\alpha $
is a free parameter. This coset space has been studied in Ref. \cite{12} and
interpreted as a closed inhomogeneous universe \cite{11'}.
 It can also be obtained by
an $O(2,2,I\hspace{-.15cm}R)$ rotation of a product of two dimensional
Lorentzian and Euclidean black holes \cite{13}, \cite{14}. Parametrizing $%
g_2 $ as
$$
g_2=\exp (i\frac{\rho +\lambda }{\sqrt{2}}\sigma _2)\exp (i\theta \sigma
_3)\exp (i\frac{\rho -\lambda }{\sqrt{2}}\sigma _2)
$$
and gauge fixing by choosing
$$
g_1=\left( \matrix{\cos\psi&\sin\psi\cr
-\sin\psi&\cos\psi\cr}\right) ,
$$
one finds that the metric is given by
\begin{equation}
ds^2=-d\psi ^2+d\theta ^2+G_{\rho \rho }d\rho ^2+G_{\lambda \lambda
}d\lambda ^2,\nonumber
\end{equation}
where
\begin{eqnarray}
G_{\rho\rho}&=&\frac{4\cos^2\psi\cos^2\theta(1+\sin\alpha)}
{(1-\cos2\psi\cos2\theta)+\sin\alpha(\cos2\psi-\cos2\theta)},\\
G_{\lambda\lambda}&=&\frac{4\sin^2\psi\sin^2\theta(1-\sin\alpha)}
{(1-\cos2\psi\cos2\theta)+\sin\alpha(\cos2\psi-\cos2\theta)},
\nonumber\end{eqnarray}
The antisymmetric tensor and the dilaton field are found to be
\begin{eqnarray}
B_{\rho\lambda}&=&
\frac{\cos2\psi-\cos2\theta+\sin\alpha(1-\cos2\psi\cos2\theta)}
{(1-\cos2\psi\cos2\theta)+\sin\alpha(\cos2\psi-\cos2\theta)},\\
\Phi&=&-\frac{1}{2}\ln[1-
\cos2\psi\cos2\theta+\sin\alpha(\cos2\psi-\cos2\theta)],
\end{eqnarray}
In terms of advanced and retarded coordinates
\begin{eqnarray}
u=\frac{1}{2}(\psi+\theta)-\frac{\pi}{4}&,&
v=\frac{1}{2}(\psi-\theta),\nonumber
\end{eqnarray}
and choosing $\alpha =0$, the metric, the antisymmetric tensor and the
dilaton turn out to be
\begin{eqnarray}
ds^2&=&-4dudv+
\frac{4\cos^2(u+v+\pi/4)\cos^2(u-v+\pi/4)}
{1-\sin[2(u+v)]\sin[2(u-v)]}d\rho^2+\nonumber \\&&
\frac{4\sin^2(u+v+\pi/4)\sin^2(u-v+\pi/4)}
{1-\sin[2(u+v)]\sin[2(u-v)]}d\lambda^2, \label{i1}\\
B_{\rho\lambda}&=&\frac{\sin[2(u-v)]-\sin[2(u+v)]}
{1-\sin[2(u+v)]\sin[2(u-v)]}, \label{i2}\\
\Phi&=&-\frac{1}{2}\ln(1-\sin[2(u+v)]\sin[2(u-v)]). \label{i3}
\end{eqnarray}
The $\pi /2$ shift in the coordinate $u$ is such that the discontinuities to
appear in the surfaces $(u=0,\,v=0)$. On the other hand, the free parameter $%
\alpha $ has been taken to be zero in order a proper matching with plane
waves to be achieved as we will see below. Continuity of the metric, the
antisymmetric tensor and the dilaton field across the surface $u=0$
specifies them in region $III$ to be
\begin{eqnarray}
ds_{III}^2&=&-4dudv+\frac{\cos^22v}{1+\sin^22v}(d\rho^2+d\lambda^2),
\label{iii1}\\
B^{III}_{\rho\lambda}&=&-\frac{2\sin2v}{1+\sin^22v},\label{iii2}\\
\Phi^{III}&=&-\frac{1}{2}\ln(1+\sin^22v),\label{iii3}
\end{eqnarray}
while at the surface $v=0$ we have
\begin{eqnarray}
ds_{II}^2&=&-4dudv+\frac{(\cos u-\sin u)^2}{(\cos u+\sin u)^2}d\rho^2+
\frac{(\cos u+\sin u)^2}{(\cos u-\sin u)^2}d\lambda^2,
\label{ii1}\\
B^{II}_{\rho\lambda}&=&0,\label{ii2}\\
\Phi^{II}&=&-\ln \cos2u.\label{ii3}
\end{eqnarray}
Finally, continuity of the metric, antisymmetric tensor and dilaton fields
across $v=0$ of Eqs. (\ref{iii1}--\ref{iii3}) and/or $u=0$ of Eqs. (\ref{ii1}%
--\ref{ii3}) gives
\begin{eqnarray}
ds_I^2&=&-4dudv+d\rho^2+d\lambda^2,\\
B^{I}_{\rho\lambda}&=&0, \\
\Phi&=&0.
\end{eqnarray}
Thus, region $I$ is flat Minkowski space-time with constant dilaton.

We observe that regions $II,\, III$ correspond to plane wave backgrounds
moving in opposite directions. After the collision, one expects
singularities to be formed as a result of the focusing effect. In fact,
region $IV$ is singular at the surface $u+v+\pi/4=0$ which is an orbifold
singularity as discussed in Ref. \cite{12}.

Let us now examine the type of discontinuities of the derivatives of the
metric $G_{\mu \nu }$, the antisymmetric tensor $B_{\mu \nu }$ and the
dilaton field $\Phi $.

\vspace{.2cm}

{\it i) \, v=0:} \\ At the surface $v=0$ one may check that the derivative
of the metric and the dilaton field are continuous, i.e.,
\begin{eqnarray}
\partial_uG^{IV}_{ij}|_{v=0}&=&\partial_uG^{II}_{ij}|_{v=0},\nonumber
\\
\partial_vG^{IV}_{ij}|_{v=0}&=&\partial_vG^{II}_{ij}|_{v=0},\nonumber\\
\partial_u\Phi^{IV}|_{v=0}&=&\partial_u\Phi^{II}|_{v=0}.
\nonumber \\
\partial_v\Phi^{IV}|_{v=0}&=&\partial_v\Phi^{II}|_{v=0}.
\end{eqnarray}
and thus a shock wave discontinuity appears in the $v=0$ surface. However,
the antisymmetric tensor is discontinuous and it is straightforward to check
that the corresponding discontinuity in the antisymmetric field strength is
of the type (\ref{H}). Indeed, one may find that
\begin{eqnarray}
H^{IV}_{u\rho\lambda}&=&
\frac{2(\cos[2(u-v)]-\cos[2(u+v)])}{1-\sin[2(u+v)]
\sin[2(u-v)]} +\nonumber \\ &&
2\sin4u\frac{\sin[2(u-v)]-\sin[2(u+v)]}{(1-\sin[2(u+v)]
\sin[2(u-v)])^2},\\
H^{IV}_{v\rho\lambda}&=&
-\frac{2(\cos[2(u-v)]+\cos[2(u+v)])}{1-\sin[2(u+v)]
\sin[2(u-v)]}+\nonumber \\    &&
2\sin4v\frac{\sin[2(u-v)]-\sin[2(u+v)]}{(1-\sin[2(u+v)]
\sin[2(u-v)])^2}.
\end{eqnarray}
Thus we have
\begin{eqnarray}
H^{IV}_{u\rho\lambda}|_{v=0}&=&H^{II}_{u\rho\lambda}|_{v=0}=0,\nonumber \\
H^{IV}_{v\rho\lambda}|_{v=0}&=& \frac{4\sin2u}{\cos^22u}
,\nonumber \\H^{II}_{v\rho\lambda}|_{v=0}&=&=0
\end{eqnarray}
and the discontinuity of the antisymmetric field strength, recalling that $%
u^{\mu}=(-1/2,0,0,0)$ for the hypersurface $v=0$, is of the type (\ref{H})
with $h$ given by
$$
h=-\frac{8\sin2u}{\cos^22u} .%
$$

{\it ii) \thinspace u=0:} \\ In the surface $u=0$ we have
\begin{eqnarray}
H^{IV}_{u\rho\lambda}|_{u=0}=H^{III}_{u\rho\lambda}&=&0, \nonumber\\
H^{IV}_{v\rho\lambda}|_{u=0}=H^{III}_{v\rho\lambda}|_{u=0}&=&
 -\frac{4\cos2v}{1+\sin^22v}-\frac{4\sin2v\sin4v}{(1+\sin^22v)^2}
,\end{eqnarray}
so that the antisymmetric field strength is continuous across $u=0$.
Furthermore, we find that
\begin{eqnarray}
\partial_vG^{IV}_{ij}|_{u=0}&=&\partial_vG^{III}_{ij}|_{u=0},\nonumber
\\
\partial_v\Phi^{IV}|_{u=0}&=&\partial_v\Phi^{III}|_{u=0},
\nonumber \\
\partial_u\Phi^{IV}|_{u=0}&=&\partial_u\Phi^{III}|_{u=0},
\end{eqnarray}
However, there exist a discontinuity in the derivative of the metric
\begin{eqnarray}
\partial_uG^{IV}_{\rho\rho}&=& -\frac{2\cos2v}{1+\sin^22v}, \nonumber \\
\partial_uG^{IV}_{\lambda\lambda}&=&\frac{2\cos2v}{1+\sin^22v},\label{dis}
\end{eqnarray}
while
\begin{eqnarray}
\partial_uG^{III}_{\rho\rho}=0&,&\partial_uG^{III}_{\lambda\lambda}=0.
\label{dis1}
\end{eqnarray}
Comparing Eqs. (\ref{dism},\ref{dis},\ref{dis1}) we find that
\begin{equation}
\gamma _{\lambda \lambda }=-\gamma _{\rho \rho }=\frac{2\cos 2v}{1+\sin ^22v}
\end{equation}
and one may check that the condition (\ref{BS}) is indeed satisfied.

Let us note also that the backgrounds (\ref{iii1}--\ref{iii3}) and (\ref{ii1}%
--\ref{ii3}) have a coset CFT interpretation. To see this, one may check
that these backgrounds are singular limits of the $SL(2,I\hspace{-.15cm}%
R)\!\times SU(2)/I\hspace{-.15cm}R\!\times I\hspace{-.15cm}R$ model.
Introducing the parameters $\varepsilon, \, \varepsilon^{\prime}$ and
rescaling the coordinates and the coupling as
\begin{equation}
u\rightarrow \varepsilon u, \hspace{.5cm} v\rightarrow
\varepsilon^{\prime}v, \hspace{.5cm} \rho\rightarrow \sqrt{%
\varepsilon\varepsilon^{\prime}}\rho, \hspace{.5cm} \lambda\rightarrow \sqrt{%
\varepsilon\varepsilon^{\prime}} \lambda, \hspace{.5cm} \alpha^{\prime}%
\rightarrow \varepsilon\varepsilon^{\prime} \alpha^{\prime},
\end{equation}
the background (\ref{iii1}--\ref{iii3}) in region $III$ can be obtained in
the limit $\varepsilon \rightarrow 0$, the background (\ref{ii1}--\ref{ii3})
in region $II$ in the $\varepsilon^{\prime}\rightarrow 0$ limit and the flat
Minkowski space in $I$ when $\varepsilon\rightarrow 0$ and $%
\varepsilon^{\prime}\rightarrow 0$.

We have discussed here the conditions which must be satisfied in order the
beta-function equations to not be affected by the presence of
discontinuities in the target space. We found that these discontinuities
must be across null hypersurfaces and thus gravitational wave backgrounds,
shock or impulsive, are exact in string theory. As a result, backgrounds
resulting from linear superpositions of independent shock or impulsive waves
with parallel propagating wave-fronts are exact as well. We have also
discussed the $SL(2,I\hspace{-.15cm}R)\!\times \!SU(2)/I\hspace{-.15cm}%
R\!\times \!I\hspace{-.15cm}R$ WZW model and we have shown that this model
can be considered as the resulting space-time of the collision of two plane
waves. It should be noted, however, that the above leading order solution,
indicates what one should expect from a collision of plane waves. Finally,
in the present framework, an interesting possibility is the construction of
cosmological models built up from gravitational waves \cite{G} where
space-time singularities can be understood as a focusing result.

\vspace{1cm}

We would like to thank C. Bachas for discussions.

\end{document}